\documentclass[a4paper,fleqn,usenatbib]{mnras}

\usepackage{newtxtext}
\usepackage{newtxmath}

\usepackage[T1]{fontenc}
\usepackage{ae,aecompl}

\usepackage{graphicx}	
\usepackage{amsmath}	
\usepackage{amssymb}	

\usepackage{threeparttable}
\usepackage{multirow}
\usepackage{booktabs}

\usepackage{color}

\newcommand{\vv}{\upsilon}

\def\C#1{#1}

\title[Cosmic ray driven galactic winds]{The dependence of cosmic ray driven galactic winds on halo mass}

\author[Jacob et al.]{
Svenja Jacob,$^{1, 2}$\thanks{E-mail: svenja.jacob@h-its.org}
R\"udiger Pakmor,$^{1}$
Christine M. Simpson,$^{1}$
Volker Springel,$^{1,2,3}$
\newauthor{
Christoph Pfrommer$^{4,1}$ \vspace*{0.2cm}}
\\
$^{1}$Heidelberg Institute for Theoretical Studies, Schloss-Wolfsbrunnenweg 35, D-69118 Heidelberg, Germany\\
$^{2}$Zentrum f\"ur Astronomie der Universit\"at Heidelberg, ARI, M\"onchhofsstrasse 12-14, D-69120 Heidelberg, Germany\\
$^{3}$Max-Planck-Institut f\"ur Astrophysik, Karl-Schwarzschild-Str. 1, 85741 Garching, Germany\\
$^{4}$Leibniz-Institut f\"ur Astrophysik Potsdam (AIP), An der Sternwarte 16, D-14482 Potsdam, Germany
}

\date{Accepted XXX. Received YYY; in original form ZZZ}

\pubyear{2017}

\begin{document}
\label{firstpage}
\pagerange{\pageref{firstpage}--\pageref{lastpage}}
\maketitle

\begin{abstract} 
Galactic winds regulate star formation in disk galaxies and help to enrich the circum-galactic medium. They are therefore crucial for galaxy formation, but their driving mechanism is still poorly understood.  Recent studies have demonstrated that cosmic rays (CRs) can drive outflows if active CR transport is taken into account. Using hydrodynamical simulations of isolated galaxies with virial masses between $10^{10}$ and $10^{13}\rmn{~M_\odot}$, we study how the properties of CR-driven winds depend on halo mass. CRs are treated in a two-fluid approximation and their transport is modelled through isotropic or anisotropic diffusion.  We find that CRs are only able to drive mass-loaded winds beyond the virial radius in haloes with masses below $10^{12}\rmn{~M_\odot}$. For our lowest examined halo mass, the wind is roughly spherical and has velocities of $\sim20\rmn{~km\;s^{-1}}$. With increasing halo mass, the wind becomes biconical and can reach ten times higher velocities.  The mass loading factor drops rapidly with virial mass, a dependence that approximately follows a power-law with a slope between $-1$ and $-2$. This scaling is slightly steeper than observational inferences, and also steeper than commonly used prescriptions for wind feedback in cosmological simulations.  The slope is quite robust to variations of the CR injection efficiency or the CR diffusion coefficient.  In contrast to the mass loading, the energy loading shows no significant dependence on halo mass. While these scalings are close to successful heuristic models of wind feedback, the CR-driven winds in our present models are not yet powerful enough to fully account for the required feedback strength.
 \end{abstract}

\begin{keywords}
galaxies: formation, starburst -- hydrodynamics -- cosmic rays
\end{keywords}



\section{Introduction}

Galactic winds play an important role in the formation and evolution of galaxies. Observations demonstrate that they are common at higher redshifts as well as in star bursting galaxies in the local Universe \citep[see][for reviews]{Veilleux2005, Alsabti2017}.  Galactic winds might be able to transport chemically enriched material from the star forming disk to the circum-galactic medium and help to explain the observed metal abundances there \citep{Aguirre2001, Oppenheimer2006, Oppenheimer2008, Booth2012, Tumlinson2011}. Moreover, the wind material is at least temporarily not available for star formation in the disk.

The last point, in particular, makes galactic winds crucial in simulations of galaxy formation that typically suffer from an overproduction of stars unless very strong feedback models are invoked \citep[e.g.][]{Springel2003, Stinson2013, Puchwein2013}. So far, most simulations, especially on cosmological scales, employ empirical models to drive winds. Those sometimes include the creation of special wind particles that either cannot cool or are temporarily decoupled from hydrodynamics \citep[for a review]{Somerville2015}.

To improve the subgrid prescriptions, a better knowledge of the physical driving mechanisms of the winds is essential. Most models are based on the notion that some aspect of stellar feedback drives the winds, but it remains unclear which part of the feedback physics launches the outflow. One possibility is the direct thermal and mechanical energy input from (several) supernovae (SNe) \citep{Dekel1986, Efstathiou2000, Creasey2013, Martizzi2016, Girichidis2016b}.  Furthermore, the radiation pressure from young stars might be able to accelerate the gas, although the required opacity is still subject of debate \citep{Murray2005, Krumholz2012, Hopkins2012, Rosdahl2015, Skinner2015}.

Another possibility to drive galactic winds are cosmic rays (CRs) \citep{Ipavich1975, Breitschwerdt1991, Zirakashvili1996, Breitschwerdt2002, Everett2008, Samui2010,  Recchia2016}. CRs are relativistic particles that permeate the ISM with an energy density that is comparable to the kinetic energy density and the energy density in magnetic fields \citep{Boulares1990}.
CRs interact with the thermal gas via magnetic fields, which leads to an additional, effective pressure \citep[see][for reviews]{Strong2007, Zweibel2013}. Therefore, gradients in the CR pressure exert a force on the gas.

However, CRs can only efficiently drive winds if they can move relative to the thermal gas. In this case, extended CR pressure gradients form above and below the disk which can then accelerate the gas \citep{Salem2014a}. The required transport mechanism depends on the detailed physics of CR propagation. It can be modelled as either diffusion \citep{Salem2014a, Pakmor2016wind} or streaming \citep{Uhlig2012, Ruszkowski2017, Wiener2017}. (Magneto-) hydrodynamic simulations have demonstrated that at least one of these transport mechanisms is required to produce galactic outflows \citep{Jubelgas2008, Uhlig2012, Booth2013, Hanasz2013, Salem2014a, Salem2014b, Salem2016, Pakmor2016wind, Simpson2016, Girichidis2016a, Ruszkowski2017, Wiener2017}.

Still, a driving mechanism that generates an outflow in one galaxy might create winds with vastly different properties in a galaxy with higher or lower mass, or not drive an outflow at all. Observations of nearby starburst galaxies give some indication how certain wind properties depend on galaxy mass despite large uncertainties \citep{Heckman2015,Chisholm2017}. In simulations, this question has mostly been studied for SN-driven winds \citep{Creasey2013, Muratov2015, Li2017, Fielding2017}. To our knowledge, previous works on CR-driven winds with diffusive CRs have not focused on the halo mass dependence of the wind and mostly considered only a single, or at most two, halo masses. The halo mass dependence of winds that are driven by streaming CRs has been analysed by \citet{Uhlig2012} with three halo masses.

In this work, we study in detail which galaxies can produce CR-driven winds and how the wind properties depend on halo mass. To this end, we simulate a set of idealized, isolated galaxies that include CR diffusion, similar to the setup in \citet{Pakmor2016wind}. We vary the virial mass of the galaxy between $10^{10}$ and $10^{13}\;\rmn{M_\odot}$ and test different aspects of CR physics, such as isotropic and anisotropic diffusion.  Moreover, we compare our results to observations and empirical wind models.

This paper is structured as follows. We introduce our simulations in Section~\ref{sec:sims} and present the results in Section~\ref{sec:results}. In Section~\ref{sec:wind_dev}, we show qualitatively the formation of winds, and in Section~\ref{sec:wind_prop} we quantify the star formation efficiency, the mass loading and the energy loading in relation to the halo mass. We discuss further aspects of the simulations in Section~\ref{sec:discussion} and conclude in Section~\ref{sec:conclusion}.


\section{Simulations}
\label{sec:sims}

We simulate a set of isolated galaxies to analyse the formation and properties of CR-driven winds. This allows us to cleanly focus on the effect of halo mass, and on differences caused by certain other aspects of CR physics.

\subsection{The code}

We use the moving-mesh code {\small AREPO} \citep{Springel2010} with an improved second-order scheme \citep{Pakmor2016arepo} to solve the magnetohydrodynamical equations, \C{which are coupled to an equation for the CR energy density \citep{Pfrommer2017}. Hence, we solve the equations for mass, momentum and energy conservation together with evolution equations for the CR energy density and the magnetic field as given by 
\begin{align}
\frac{\partial \rho}{\partial t} + \bmath{\nabla \cdot} \left(\rho \bmath{\vv} \right) &= 0\\
\frac{\partial \left(\rho \bmath{\vv}\right)}{\partial t} + \bmath{\nabla \cdot} \left[ \rho \bmath{\vv} \bmath{\vv}^\rmn{T} +P \bmath{I} - \bmath{B} \bmath{B}^\rmn{T} \right] &= - \rho \nabla \Phi\\
\frac{\partial \varepsilon }{\partial t} + \bmath{\nabla \cdot} \left[(\varepsilon + P) \bmath{\vv} - \bmath{B} \left( \bmath{\vv \cdot B}\right) \right] &= P_\rmn{cr} \bmath{\nabla \cdot \vv} + \Lambda_\rmn{th} + \Gamma_\rmn{th} \\
\frac{\partial \varepsilon_\rmn{cr}}{\partial t} + \bmath{\nabla \cdot} \left[\varepsilon_\rmn{cr} \bmath{\vv} - \kappa_\rmn{cr} \bmath{\hat{b}} \left(\bmath{\hat{b} \cdot \nabla} \varepsilon_\rmn{cr} \right)\right] &= - P_\rmn{cr} \bmath{\nabla \cdot \vv} + \Lambda_\rmn{cr} + \Gamma_\rmn{cr}\\
\frac{\partial \bmath{B}}{\partial t} + \bmath{\nabla \cdot} \left[\bmath{B \vv}^\rmn{T} - \bmath{\vv B}^\rmn{T} \right] &= 0.
\end{align}
Here, $\rho$ denotes the gas density, $\bmath{\vv}$ the gas velocity and $\bmath{B}$ the magnetic field strength. $P$ is the total pressure with contributions from the thermal gas, the CRs and the magnetic field,
\begin{equation}
P = P_\rmn{th} + P_\rmn{cr} + \frac{\bmath{B}^2}{2}.
\end{equation}
$\varepsilon_\rmn{cr}$ is the CR energy density, whereas $\varepsilon$ is the total energy density without CRs,
\begin{equation}
\varepsilon = \varepsilon_\rmn{th} + \frac{\rho \bmath{\vv}^2}{2} + \frac{\bmath{B}^2}{2}.
\end{equation}
Thermal energy density and thermal pressure are related by an equation of state, $P_\rmn{th} = (\gamma_\rmn{th} -1) \varepsilon_\rmn{th}$, with the adiabatic index $\gamma_\rmn{th}$. Similarly, CR energy and CR pressure are related by  $P_\rmn{cr} = (\gamma_\rmn{cr} -1) \varepsilon_\rmn{cr}$, where $\gamma_\rmn{cr}$ is an effective adiabatic index for the CRs. $\bmath{I}$ describes the identity matrix and $\phi$ is the gravitational potential. $\Lambda_\rmn{th}$ and $\Gamma_\rmn{th}$ describe gain and loss terms for the thermal gas and $\Lambda_\rmn{cr}$ and $\Gamma_\rmn{cr}$ describe gain and loss terms for the CRs. We take into account CR diffusion along magnetic field lines. The CR diffusion coefficient is denoted by $\kappa_\rmn{cr}$ and the unit vector pointing in the direction of the magnetic field is denoted by $\bmath{\hat{b}} = \bmath{B} / |\bmath{B}|$. We neglect CR streaming and CR Alfvén wave losses for simplicity (see Wiener et al. 2017, for a discussion of these effects in comparison to CR diffusion).
}

 {\small AREPO} uses a Voronoi tessellation to discretise space with a refinement scheme that keeps the mass in all cells approximately constant. Additionally, we apply an upper limit on the volume of a cell and we only allow a factor of ten difference in the volume between adjacent cells to make the mesh resolution vary more smoothly.  In all our simulations, we take the self-gravity of the gas and stars into account based on a tree-based gravity solver. However, the dark matter halo is described by a static background potential in our fiducial standard simulations, except for a subset of our simulations where we also model the dark matter explicitly.

We employ the cooling and star formation prescriptions of \citet{Springel2003} with an effective equation of state. But importantly, we do not include an empirical wind model in our simulations.  We use the CR two-fluid model introduced for {\small AREPO} by \citet{Pfrommer2017} and treat CRs as an additional fluid with an adiabatic index of $\gamma_\rmn{cr} = 4/3$, corresponding to the ultra-relativistic limit. We apply the subgrid model for the acceleration of CRs at supernova remnants, where we inject $10^{48}\;\rmn{erg}$ of CR energy per solar mass of star formation. Moreover, we take into account that CRs lose energy to the thermal gas due to Coulomb and hadronic interactions.

CRs are always advected with the thermal gas. But additional transport mechanisms, such as diffusion or streaming, can introduce a relative motion between gas and CRs. In this paper, we study the effects of isotropic and anisotropic diffusion and leave the analysis of streaming to future work. To this end we use the diffusion solver from \citet{Pakmor2016code} that also allows anisotropic diffusion if a magnetic field is present.  In our fiducial simulations with isotropic diffusion, we use a constant diffusion coefficient of $\kappa_\rmn{cr} = 10^{28}\;\rmn{cm^2\;s^{-1}}$. For anisotropic diffusion, we use the same value for diffusion along magnetic field lines and set the perpendicular diffusion coefficient to zero. This results in a lower `effective' diffusion coefficient for anisotropic diffusion if the magnetic fields are fully tangled.

In the simulations with anisotropic diffusion, we also include magnetic fields. For this purpose, we use the ideal MHD module for {\small AREPO} from \citet{Pakmor2011} with the Powell scheme for divergence cleaning \citep{Powell1999}. The magnetic field is initially uniform and oriented along the $\hat{\bmath{x}}$-direction, with a field strength of $10^{-10}\rmn{~G}$.

\subsection{The simulation setup}
\label{sec:setup}

\begin{table}
\caption{Galaxy properties.}
\label{tab:masses}
\setlength{\tabcolsep}{2pt}
\begin{threeparttable}
\begin{tabular}{l r r r r r r}
\toprule
Halo & \multicolumn{1}{c}{$\vv_\rmn{vir}$} & \multicolumn{1}{c}{$M_\rmn{vir}$} &\multicolumn{1}{c}{$r_\rmn{vir}$}& \multicolumn{1}{c}{$L$\tnote{(1)}} & \multicolumn{1}{c}{$M_\rmn{ref}$\tnote{(2)}}& \multicolumn{1}{c}{$r_\rmn{cyl}$\tnote{(3)}}\\
	&\multicolumn{1}{c}{$\rmn{(km\;s^{-1})}$}	&\multicolumn{1}{c}{$(\rmn{M_\odot})$}	&\multicolumn{1}{c}{$(\rmn{kpc})$}&\multicolumn{1}{c}{$(\rmn{Mpc})$} & \multicolumn{1}{c}{$(\rmn{M_\odot})$}& \multicolumn{1}{c}{$(\rmn{kpc})$}\\
\midrule
Halo 10.0 		&$35$		&$1.00\times10^{10}$	&$35$				&$1.0$	& $1.43\times 10^{3}$	&$10$ \\
Halo 10.5		&$55$		&$3.87\times10^{10}$	&$55$				&$1.5$	& $5.80\times 10^{3}$	&$15$\\
Halo 11.0		&$75$		&$9.81\times10^{10}$	&$75$				&$1.5$	& $1.43\times 10^{4}$	&$20$\\
Halo 11.5		&$110$		&$3.09\times10^{11}$	&$110$		&$1.5$	& $4.64\times 10^{4}$	&$25$ \\
Halo 12.0		&$160$		&$9.52\times10^{11}$	&$160$		&$1.5$	& $1.43\times 10^{5}$	&$30$\\
Halo 13.0		&$340$		&$9.14\times10^{12}$	&$340$		&$2.5$	& $1.43\times 10^{6}$	&$40$\\
\bottomrule
\end{tabular}
\begin{tablenotes}
\item[(1)] Boxsize
\item[(2)] Reference gas mass for refinement and derefinement 
\item[(3)] Radius of the cylinder used to calculate the SFR and the mass in the disk
\end{tablenotes}
\end{threeparttable}
\end{table}

\begin{table*}
\caption{Physics models.}
\label{tab:models}
\begin{threeparttable}
\begin{tabular}{l c c c c c c c c c}
\toprule
Model & CRs  & CR diff & CR & MHD & empirical & dm & N\tnote{(2)} & $\varepsilon_\rmn{cr}\tnote{(3)}$ & $\kappa_\rmn{cr}$\tnote{(4)}\\
&&&\multicolumn{1}{c}{cooling}&&\multicolumn{1}{c}{model\tnote{(1)}}& \multicolumn{1}{c}{halo} && $\rmn{(erg\;M_\odot^{-1})}$ & $\rmn{(cm^2\;s^{-1})}$\\
\midrule
CRs, iso diff		& yes& iso 	&yes	& no & no & static &  $1.0\times10^6$	&$1\times10^{48}$ & $1 \times 10^{28}$\\
CRs, aniso. diff 	& yes& aniso&yes	 & yes & no & static &  $1.0\times10^6$	&$1\times10^{48}$ & $1 \times 10^{28}$\\
no CRs 			& no & no 	&no 	& no & no & static & $1.0\times10^6$	& -- & --\\
CRs				& yes & no 	&yes	& no& no& static &  $1.0\times10^6$		&$1\times10^{48}$ & $1 \times 10^{28}$\\
low efficiency	& yes& iso 	&yes	& no & no & static &  $1.0\times10^6$	&$3\times10^{47}$ & $1 \times 10^{28}$\\
high efficiency	& yes& iso 	&yes	& no & no & static &  $1.0\times10^6$	&$3\times10^{48}$ & $1 \times 10^{28}$\\
low kappa		& yes& iso 	&yes	& no & no & static &  $1.0\times10^6$	&$1\times10^{48}$ & $3 \times 10^{27}$\\
high kappa		& yes& iso 	&yes	& no & no & static &  $1.0\times10^6$	&$1\times10^{48}$ & $3 \times 10^{28}$\\
very high kappa	& yes& iso 	&yes	& no & no & static &  $1.0\times10^6$	&$1\times10^{48}$ & $1 \times 10^{29}$\\
no cooling		&yes & iso	&no		&no	 & no & static & $1.0\times 10^6$	&$1\times10^{48}$ & $1 \times 10^{28}$\\
live DM			& yes& iso 	&yes	& no & no & live &  $1.5\times10^5$		&$1\times10^{48}$ & $1 \times 10^{28}$\\
low resolution			& yes& iso 	&yes	& no & no & static & $1.5\times10^5$	&$1\times10^{48}$ & $1 \times 10^{28}$\\
high resolution			& yes& iso 	&yes	& no & no & static & $5.0\times10^6$	&$1\times10^{48}$ & $1 \times 10^{28}$\\
empiric model	& no & no 	&no		& no & yes & live &  $1.0\times10^6$ 	& -- & --\\
\bottomrule
\end{tabular}
\begin{tablenotes}
\item[(1)] From \citet{Vogelsberger2013}
\item[(2)] Initial number of cells 
\item[(3)] CR injection efficiency in erg per solar mass of star formation
\item[(4)] CR diffusion coefficient
\end{tablenotes}
\end{threeparttable}
\end{table*}

Our initial conditions consist of a rotating gas sphere in a static Hernquist potential \citep{Hernquist1990}. We create the gas sphere as in \citet{Springel2003} with an angular momentum distribution that is derived from the fitting formula of \citet{Bullock2001}. \C{Due to the rotation, the gas is not in perfect hydrostatic equilibrium and a small amount of gas is initially unbound. This results in some artificial mass loss at the beginning of the simulation (see also third panel of Fig.~\ref{fig:load_time}).}
The galaxies in our sample have virial velocities between $\vv_\rmn{vir} = 35\rmn{~km\;s^{-1}}$ and $\vv_\rmn{vir} = 340\rmn{~km\;s^{-1}}$ (for details see Table~\ref{tab:masses}). The corresponding virial masses, $M_\rmn{vir} = \vv_\rmn{vir}^3/(10 G H_\rmn{0})$, range between $10^{10}$ and $10^{13}h^{-1}\rmn{M_\odot}$\footnote{Here, $H_0 = 100 h \rmn{~ km\;s^{-1}\;Mpc^{-1}}$ is the Hubble constant and $G$ the gravitational constant. For the remaining part of the paper, we set $h=1$, which corresponds to redshift $0.7$.}. 
We use the exponent of the halo mass to label the simulations, for example ``Halo~10.0'' for simulations of the halo with the lowest mass.
The scale radius of the Hernquist potential is chosen such that the concentration of an equivalent NFW profile would be $5$. The spin parameter for the gas is $0.05$ and the gas fraction is $0.15$. We keep these parameters the same for all halo masses in order to preserve self-similarity.

The haloes are placed in a box that is large enough to capture the full evolution of the wind and thus depends on the mass of the galaxy. The box sizes vary between $1$ and $2.5\;\rmn{Mpc}$ and are listed for individual simulations in Table~\ref{tab:masses}. The maximal allowed cell volume changes with box size and ranges between $1.5\times10^{5}$ and $2.3 \times 10^6\rmn{~kpc^3}$. We start all simulations with $10^6$ cells in the halo, which implies a decreasing mass resolution with increasing galaxy mass. For each halo mass, the reference mass that is targeted by the refinement scheme is listed in Table~\ref{tab:masses}.

We evolve the initial conditions for $6\;\rmn{Gyrs}$, such that winds can also develop late in the galaxy's evolution. In all simulations, the gas is allowed to cool radiatively and form stars but the included aspects of CR physics vary. Our fiducial runs contain CRs with isotropic diffusion. Here, the effect of CRs is strongest and not intertwined with the effects of the magnetic field. We also repeat our simulations with anisotropic diffusion to study its impact on the wind.

Furthermore, we carry out a number of additional reference runs. First, we repeat the simulations without CRs. We then include CRs but only advect them with the gas. In order to test the robustness of our results, we run simulations with different values for the CR injection efficiency, $\varepsilon_\rmn{cr}$, and for the CR diffusion coefficient, $\kappa_\rmn{cr}$. Moreover, we analyse runs without CR cooling, with a live dark matter halo and with higher and lower resolution. As we also want to compare CR-driven winds with the winds from empirical simulation models, we run simulations with the wind model from \citet{Vogelsberger2013}, for comparison. It creates wind particles as part of the SN feedback and allows them to escape the dense, star forming gas. To this end, the hydrodynamics of the wind particles is switched off until the wind particles recouple with the gas. Table~\ref{tab:models} gives an overview of the physical models that are used in the different simulations.


\section{Results}
\label{sec:results}

We first determine in which galaxies CR-driven outflows develop and how their velocity profiles look. We then focus on mass-loaded winds and on the dependence of the wind properties on halo mass.

\subsection{Development of outflows}
\label{sec:wind_dev}

\begin{figure*}
\includegraphics{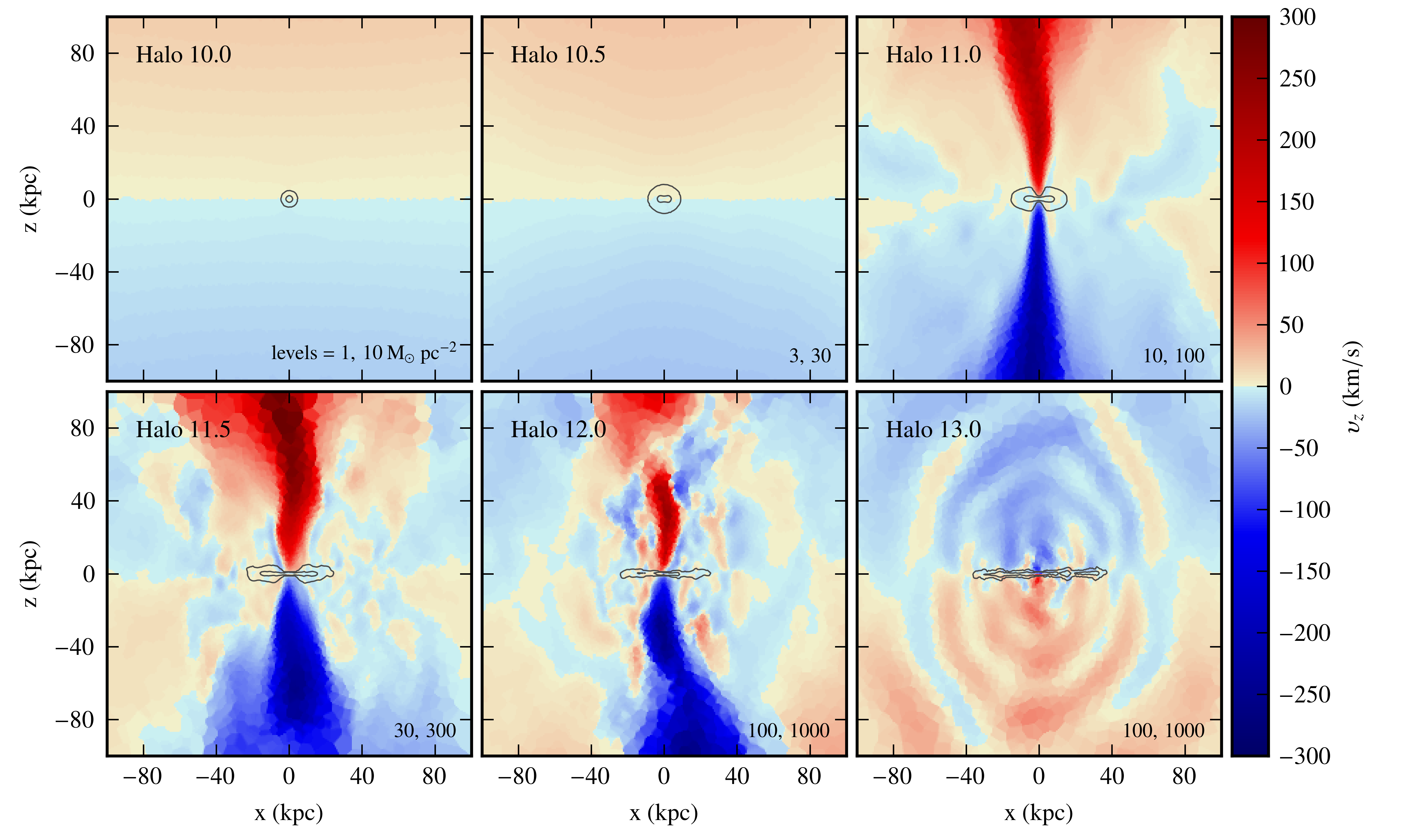}

\caption{Projections of the $z$-component of the velocity after $6\;\rmn{Gyrs}$ in our fiducial runs with isotropic CR diffusion. The mass of the displayed galaxies increases from left to right and top to bottom. The contours show the gas surface density in the central $4\rmn{~kpc}$ of the galaxy. The numbers in the bottom-right corners specify the contour levels in units of $\rmn{M_\odot\;pc^{-2}}$. The two galaxies with the lowest masses show slow, unstructured outflows whereas the galaxies with intermediate masses develop biconical outflows with higher velocities. If the mass of the galaxy is further increased, the outflow becomes weaker and is completely absent in the highest mass galaxy.}
\label{fig:projections}
\end{figure*}

For the remaining part of this paper, we refer to all material that moves away from the centre of the galaxy as `outflowing' or as an `outflow'. We later define a `wind' as an outflow that leads to a net mass loss from a cylinder around the galaxy. The cylinder's radius varies with halo mass between $10$ and $40\rmn{~kpc}$ and \C{its total height equals twice the virial radius} (see Section~\ref{sec:timeevol_loadings} for a detailed discussion of the mass loss). So a galaxy that develops an outflow does not necessarily drive a wind according to our definition.

We concentrate our analysis on the fiducial runs in which CR diffusion is isotropic. Fig.~\ref{fig:projections} shows the formation of outflows in terms of edge-on views of the velocity structure of all haloes after $6\;\rmn{Gyrs}$.
For each image, we first make a projection along the $\hat{\bmath{y}}$-direction of the mass flux perpendicular to the disk, $\rho \vv_z$. Then, we divide the result by the corresponding surface density to obtain a typical velocity in the $\hat{\bmath{z}}$-direction. We only project over the central $4 \rmn{~kpc}$ of the galaxy in order to focus on the outflow. To give an impression of the location of the galaxy, we additionally show two contours of the gas surface density. The contour levels are indicated in units of $\rmn{M_\odot\;pc^{-2}}$ in the bottom-right corner of each panel.

Fig.~\ref{fig:projections} shows that the two lowest mass haloes develop slow outflows with velocities around $30\rmn{~km\;s^{-1}}$ but without any internal structure. In Halo~10.0, the outflow is spherical and no stellar or gas disk forms. Instead, star formation proceeds in a central clump. This is different in Halo~10.5, which has a rotating gas disk. The outflow is launched above and below the disk plane but does not show the `collimation' of the outflows in higher mass haloes. 

Strong biconical outflows with velocities of more than $200\rmn{~km\;s^{-1}}$ develop in Haloes~11.0 and 11.5. In Halo~11.5, we also observe infalling gas. This mixture of inflowing and outflowing gas becomes more pronounced in Halo~12.0, while the outflow itself weakens noticeably. It is highly asymmetric and the upper half is disturbed. The strongest part of the outflow reaches only $40\rmn{~kpc}$ after $6\rmn{~Gyr}$.

No appreciable outflow forms in the highest mass halo, Halo~13.0. Even after $6\rmn{~Gyrs}$, the gas is still falling onto the centre of the galaxy. In addition to the infall motion, the velocity map in Fig.~\ref{fig:projections} shows a wave-like pattern. It occurs when the infalling gas hits the already existing disk. \C{An analysis of the vorticity profile indicates that these waves might be gravity waves, but a dedicated study is needed to confirm this preliminary result.}

\begin{figure*}
\includegraphics[width = \textwidth]{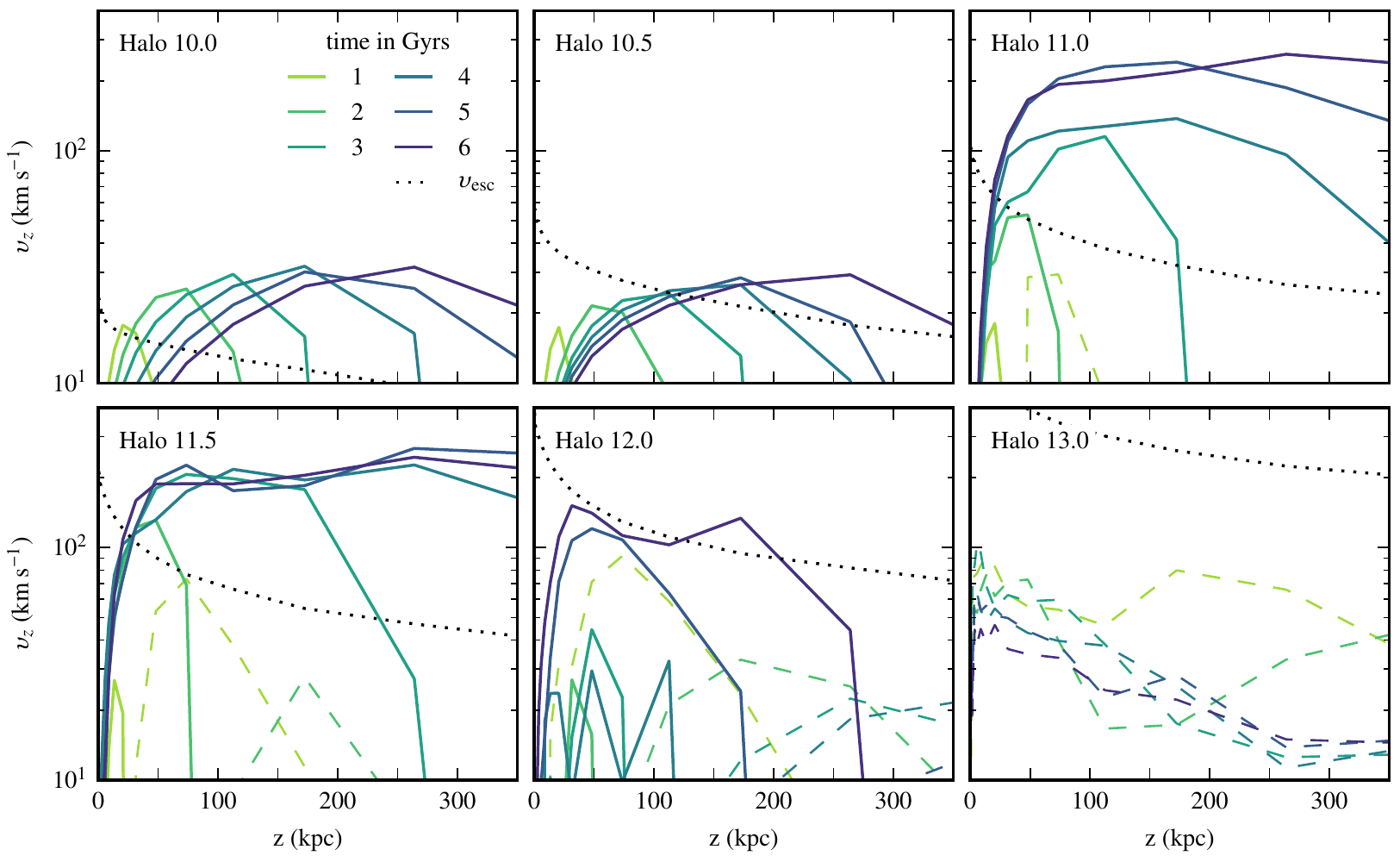}

\caption{Time evolution of the $z$-component of the gas velocity as a function of height above the disk, $z$, in our fiducial runs with isotropic CR diffusion. Solid lines indicate outflow whereas dashed lines indicate infall. The dotted lines show the escape velocity. All haloes with masses below $10^{12}\rmn{~M_\odot}$ develop strong outflows with velocities that exceed the escape speed. The maximum velocity increases with halo mass, although not linearly. In Halo~12.0, there is some outflowing material at late times but its velocity barely reaches the escape speed. No outflow develops in Halo~13.0.}
\label{fig:vel_vert}
\end{figure*}

We analyse the velocity structure of the outflows in the simulations with isotropic CR diffusion  quantitatively in Fig.~\ref{fig:vel_vert}. We consider a thin cylinder around the centre of the galaxy with a radius of $5\rmn{~kpc}$ and a total height of $1\rmn{~Mpc}$. The cylinder is sub-divided into 40 smaller cylinders stacked along the $\hat{\bmath{z}}$-direction (each with the same radius). The heights of the cylinders are logarithmically spaced from the plane of the galaxy in order to increase the resolution towards the mid plane. \C{The resolution in the wind region cannot be increased further due to the limited number of cells there.} We take a mass-weighted average of $\vv_z$ in each of the small cylinders and additionally average over the upper and lower half-planes. \C{Hence, only 20 bins are visible in Fig.~\ref{fig:vel_vert}.} Each panel in Fig.~\ref{fig:vel_vert} shows the averaged velocity profiles at six different times for a given halo mass. The solid lines indicate outflowing material whereas the dashed lines indicate infalling motion. The dotted line shows the profile of the escape velocity after $3\rmn{~Gyrs}$. It is calculated for individual cells as $\vv_\rmn{esc} = \sqrt{-2 \phi}$ from the gravitational potential $\phi$ and is averaged in the same way as the velocity.

The figure demonstrates again that distinct outflows only develop in Haloes~10.0, 10.5, 11.0 and 11.5. Here, the outflow reaches velocities that clearly exceed the escape velocity. Halo~12.0 also shows some outflowing material at late times, as can be seen in Fig.~\ref{fig:projections}, but unlike in the lower mass haloes, the velocity barely reaches the escape speed. In the most massive galaxy with $10^{13}\rmn{~M_\odot}$ (Halo~13.0),  the gas keeps infalling during the entire $6\rmn{~Gyrs}$ of our simulation.

Furthermore, Fig.~\ref{fig:vel_vert} illustrates that the outflows in Haloes 10.0 and 10.5 start to develop close to the centre after $1\rmn{~Gyr}$ and propagate outwards with time. Remarkably, the outflow is accelerated away from the centre since the maximum velocity increases for $3\rmn{~Gyrs}$ and then saturates at roughly $30\rmn{~km\;s^{-1}}$. This velocity is almost the same for the two halo masses. Close to the centre, the velocity decreases with time, which might indicate that the outflow is not replenished.

In the two haloes with higher masses, Haloes~11.0 and 11.5, the outflows also start to develop close to the centre after $1\rmn{~Gyr}$. At the same time, the outer parts of the haloes are still collapsing. This state lasts even longer in Halo~11.5. With time, the outflows propagate away from the mid plane with a velocity that is higher than in the lower mass haloes. The maximum velocity of the outflow increases rapidly and levels off at approximately $200\rmn{~km\;s^{-1}}$. Again, this value is very similar for the two haloes with biconical outflows. 
In contrast to Haloes~10.0 and 10.5, the outflows remain fast close to the centre in Haloes~11.0 and 11.5, even at late times.


\subsection{Wind properties as a function of halo mass}
\label{sec:wind_prop}

We continue our analysis with the wind properties as a function of halo mass. We first investigate how the outflows alter the star formation efficiency. We then analyse which galaxies produce mass-loaded winds and consider their mass and energy loading.

\subsubsection{Impact on star formation efficiency}

\begin{figure}
\includegraphics{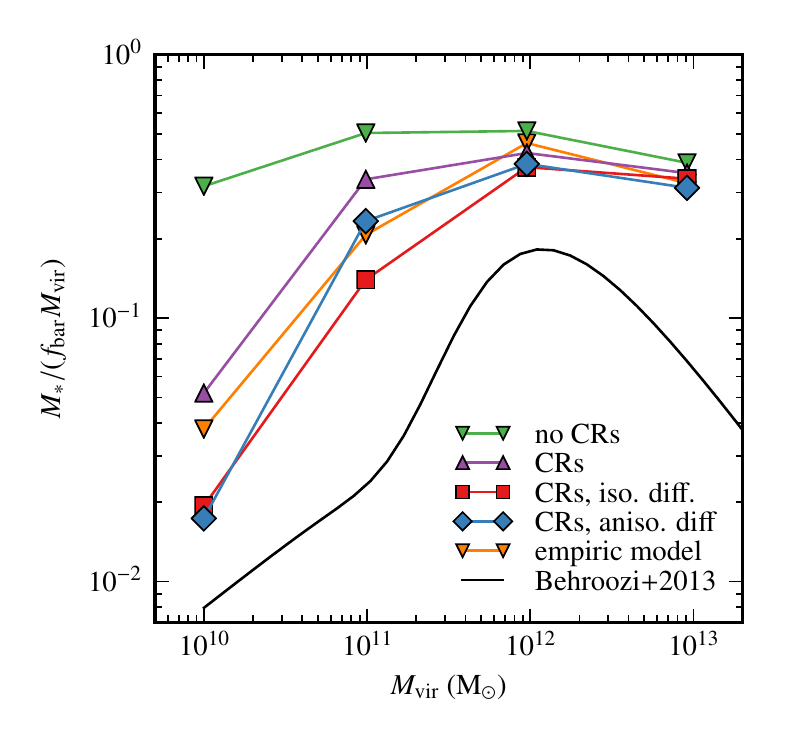}
\caption{Star formation efficiency after $6\rmn{~Gyrs}$ as a function of halo mass. The star formation efficiency is defined as the fraction of baryon mass that is converted into stars. The star formation efficiency is reduced in the presence of CRs, especially if CR-driven outflows develop in the lower mass haloes in the simulations with CR diffusion. The black line represents a fit to observational modelling from \citet{Behroozi2013b}.}
\label{fig:sfe}
\end{figure}

Galactic winds affect the efficiency of star formation within a galaxy. This property makes them an important part of simulations of galaxy evolution which is often incorporated in the form of subgrid models \citep[e.g.][for a review]{Somerville2015}. With our set of simulations, we can directly study the impact of CRs and CR-driven outflows on the amount of formed stars. Here, we define the star formation efficiency as the ratio between stellar mass, $M_\ast$, and total baryon mass in the halo, $f_\rmn{bar} M_\rmn{vir}$. The baryonic mass fraction, $f_\rmn{bar}$, is $0.15$ in all our simulations.

Fig.~\ref{fig:sfe} shows the star formation efficiency as a function of halo mass after $6\rmn{~Gyrs}$. Independent of the included physics, the star formation efficiency peaks in galaxies with $10^{11}$ to $10^{12}\rmn{~M_\odot}$. This general shape, which is already present in simulations without CRs and without an empirical wind model, is mainly a result of the cooling function. It depends on temperature and thus on halo mass. Cooling is most efficient in galaxies with masses around $\sim 10^{12}\rmn{~M_\odot}$ and hence, more stars are formed in those galaxies compared to galaxies with lower or higher masses \citep[e.g.][]{Silk1977, Rees1977}. 

The star formation efficiency is highest in the simulations with neither CRs nor an empirical wind model. After $6\rmn{~Gyrs}$, 30 to 50 per cent of the baryons are converted into stars. If CRs are included without diffusion, the star formation efficiency in Halo~10.0 drops by a factor of  $\sim 6$, from 32 to 5 per cent. In Halo~11.0, the reduction of star formation is already smaller, from 50 per cent of baryons in stars to 34 per cent. In the haloes with even higher masses, the effect of CRs decreases further. CRs reduce the star formation because they provide additional pressure support in the disk. The disk is puffed up and it becomes more difficult for the gas to collapse and form stars.

Even less stars are formed in the simulations in which outflows develop due to CR diffusion. The outflows transport material away from the disk that is then no longer available for star formation. This effect is strongest in the halo with the lowest mass, Halo~10.0. Here, the star formation efficiency drops to roughly two per cent. Hence, CR pressure support and CR-driven outflows prevent star formation almost completely and no gas or stellar disk is formed.

The effect of a galactic outflow is also noticeable in Halo~11.0. Here, only 14 per cent of the baryons are converted into stars  after $6\rmn{~Gyrs}$ if isotropic diffusion is included. This number rises to 23 per cent in the simulation with anisotropic diffusion. Thus, for this halo, it makes a difference whether isotropic or anisotropic diffusion is used. The reason is most likely the magnetic field topology. For anisotropic diffusion, CRs need field lines that are open in the vertical direction in order to diffuse out of the galaxy and drive a wind. Thus, it becomes more difficult to launch outflows and the outflows are weaker.

As shown in the previous section, no strong outflows develop in the two most massive haloes, Haloes~12.0 and 13.0. Including CR diffusion does not have a large effect on their star formation efficiency although the CR pressure support is reduced if CRs can diffuse out of the halo.

For comparison, we also show results for the simulations with the empirical wind model. The general trends are the same: the star formation efficiency peaks at $10^{12}\rmn{~M_\odot}$ and falls off to higher and lower masses. However, the empirical wind model does not shut down star formation in Halo~10.0 as efficiently as CR-driven outflows. Still, the star formation efficiency of CR-driven outflows agrees reasonably well with the empirical model. Since the wind model is very successful in more realistic, cosmological simulations \citep{Vogelsberger2013, Vogelsberger2014Illustris}, this result supports CRs as the driver of galactic winds.

We also compare the star formation efficiencies in our simulations directly with cosmological abundance matching expectations. The black line in Fig.~\ref{fig:sfe} shows a fit to the stellar mass to halo mass relation from \citet{Behroozi2013b} at redshift $0.7$.\footnote{We multiply the halo mass with the baryon fraction, $f_\rmn{b} = 0.16$ \citep{Planck2016cosmoparams}, to obtain the baryon mass as in figure 2 in \citet{Behroozi2013a}.} Overall, the star formation efficiencies are still too high, even with CRs and CR diffusion. However, the same applies for the empirical wind model, which has been tuned to reproduce the stellar mass to halo mass relation in cosmological simulations (together with AGN feedback). Hence, the simplified simulation setup as a monolithic collapse contributes substantially to the discrepancies between simulations and observational modelling in Fig.~\ref{fig:sfe}. Moreover, our simulations neglect other crucial feedback processes such as the effects from AGNs or radiation.

\subsubsection{Time evolution of mass and energy loading}
\label{sec:timeevol_loadings}

\begin{figure*}
\includegraphics[width = \textwidth]{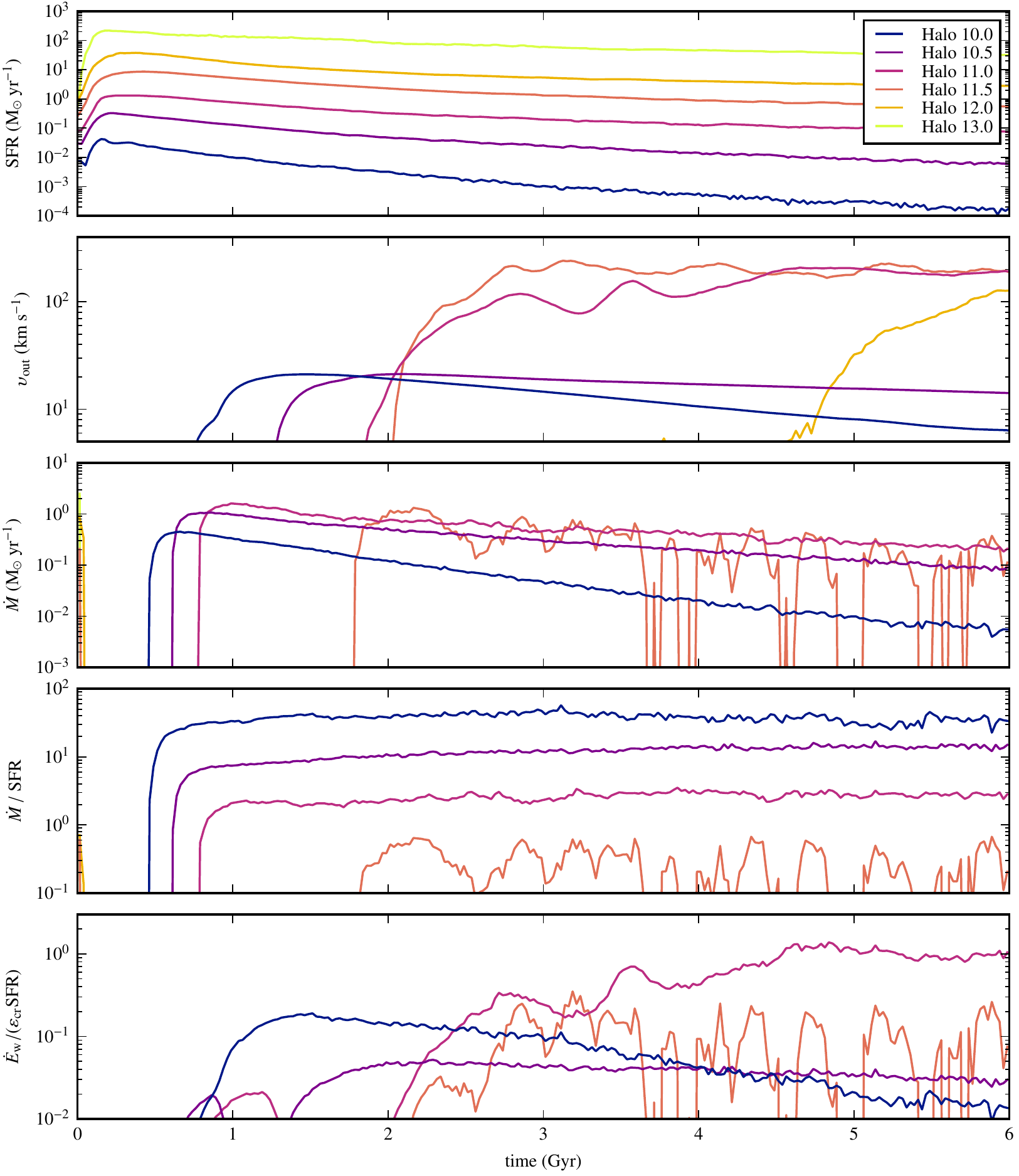}

\caption{The top three panels show SFR, outflow velocity and mass loss as a function of time for all haloes in the simulations with isotropic CR diffusion. Mass loss only occurs in the four haloes with the lowest masses and thus only those four haloes produce a wind according to our definition. The bottom two panels show mass and energy loading of the winds. The mass loading clearly scales with halo mass and is remarkably constant with time. There is no apparent scaling with halo mass of the energy loading.}

\label{fig:load_time}
\end{figure*}

\begin{figure*}
\includegraphics{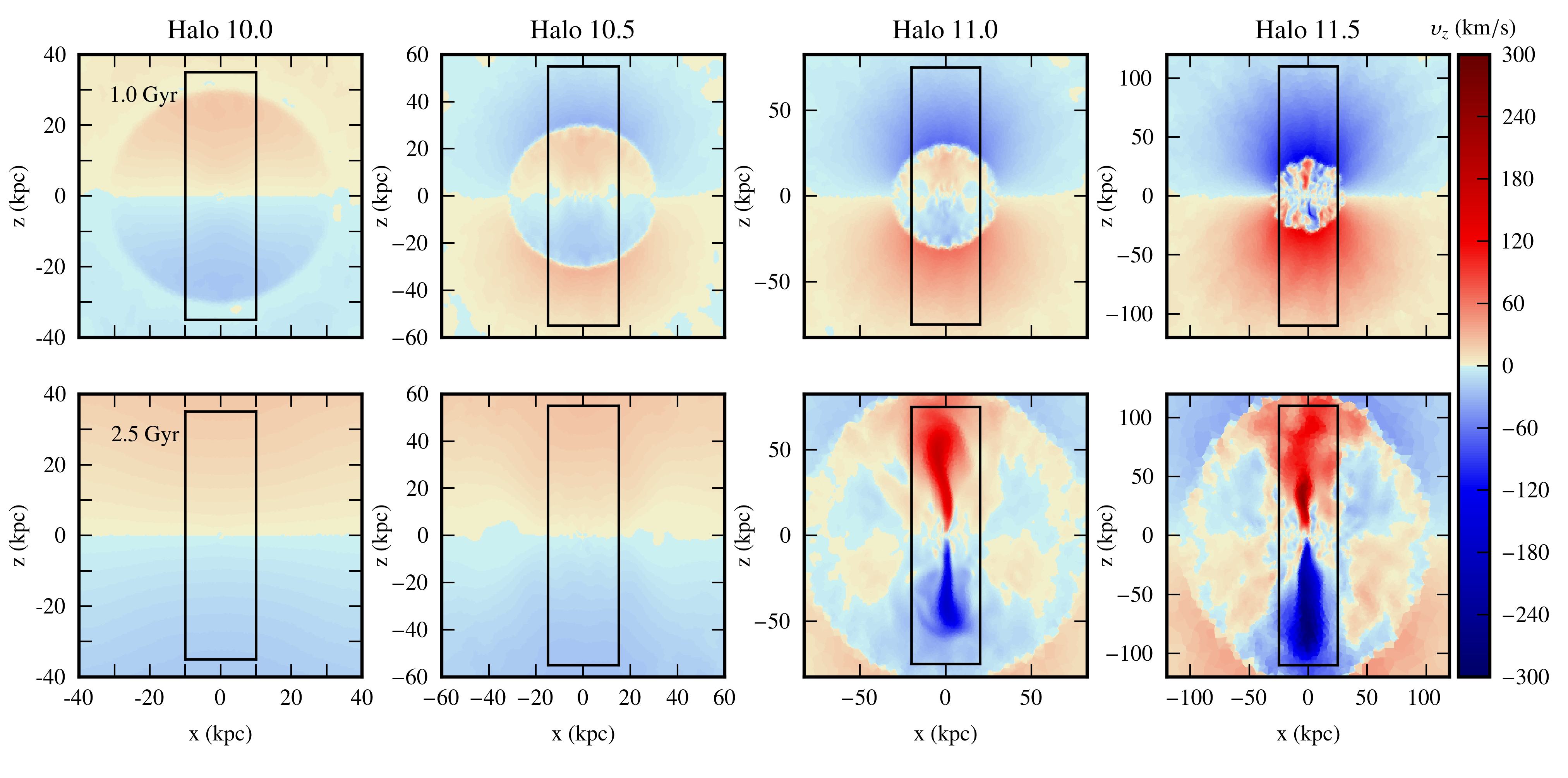}

\caption{\C{Projections of the $z$-component of the velocity after $1\rmn{~Gyr}$ (top panel) and $2.5\rmn{~Gyrs}$ (bottom panel) in our fiducial runs with isotropic CR diffusion. Each column shows one halo mass. The mass increases from left to right and the displayed region increases accordingly. The rectangular boxes demonstrate the cylinders that we use to calculate mass loss. The top panel illustrates the different onset times of the mass loss and the bottom panel shows the outflows after they have reached the virial radius in Haloes~11.0 and 11.5.}}
\label{fig:projections_boxes}
\end{figure*}

A characteristic property of galactic winds is the ratio between mass loss rate and star formation rate (SFR), which is also called the mass loading factor. Similarly, the kinetic energy in the wind can be compared to the CR energy that is injected by SNe. In this section, we discuss step-by-step how we derive these ratios for our fiducial runs with isotropic CR diffusion.

We determine the SFR in every snapshot by summing up the SFRs of individual cells within a cylinder that encloses the star forming disk. The cylinder has a height of $10\rmn{~kpc}$ for all galaxies and a radius that varies with halo mass. The radii range from $10$ to $40\rmn{~kpc}$ and are listed in Table~\ref{tab:masses}.
The resulting SFRs are shown as a function of time in the top panel of Fig.~\ref{fig:load_time}. In all simulations, there is an initial starburst which is followed by a gradual decline of the SFR. The SFR is a strong function of halo mass with a peak value of $0.05\rmn{~M_\odot\;yr^{-1}}$ in the galaxy with the lowest mass and $200\rmn{~M_\odot\;yr^{-1}}$ in the most massive galaxy.

Next, we analyse the outflow velocity at the virial radius, $r_\rmn{vir} = \vv_\rmn{vir} / (10 H_0)$ (see Table~\ref{tab:masses} for the values). We choose this radius since it scales naturally with halo mass and is far enough from the galaxy's centre so that the outflow can reach its maximum velocity.  We first determine vertical velocity profiles for each snapshot as described in Section~\ref{sec:wind_dev} and illustrated in Fig.~\ref{fig:vel_vert}. Then, we use a linear interpolation of the binned profile to obtain the outflow velocity at $r_\rmn{vir}$. We show the resulting outflow velocity as a function of time for the different halo masses in the second panel of Fig.~\ref{fig:load_time}. Here, we only show true outflow velocities and do not display infall velocities.  In the galaxy with the lowest mass, Halo~10.0, the velocity quickly reaches its maximum and then slowly declines. In contrast, the outflow velocity in the intermediate mass haloes stays roughly constant with time after the outflow reaches $r_\rmn{vir}$. While the maximum velocity in Halo~10.5 is only $20\rmn{~km\;s^{-1}}$, it reaches velocities of $200\rmn{~km\;s^{-1}}$ in Haloes~11.0 and 11.5. The time evolution of the outflow velocity shows some wiggles in Halo~11.0 and 11.5, which develop when the biconical outflows do not propagate outwards perfectly aligned with the $\bmath{\hat{z}}$-direction. So when we measure the outflow velocity, we do not always probe the centre of the biconic structure. This introduces some uncertainty to the outflow velocity and all derived quantities.  In Halo~12.0, we only observe outflowing material at $r_\rmn{vir}$ after $4.5\rmn{~Gyrs}$. As shown before, Halo~13.0 does not show any outward moving material.

A crucial quantity for characterising the outflow strength is the mass loss rate due to the outflow in relation to the star formation rate of a galaxy. Unfortunately, there is no unique way to determine the mass loading in simulations.  All results presented in this paper therefore apply to our definition of mass loss and might change for other choices.  \C{We first consider the total baryonic mass within a cylinder that is centred on the galactic centre in each snapshot. The cylinder has the same radius as the cylinder that we use to calculate the SFR (see Table~\ref{tab:masses}). In the $\bmath{\hat{z}}$-direction, the cylinder reaches the virial radius above and below the disk, such that its total height is twice the virial radius.}  Then, we determine the mass loss as $\dot{M} = \Delta M / \Delta t$ with the mass difference in the cylinder, $\Delta M$, between two consecutive snapshots that are separated by the time $\Delta t$. With this approach, we can only probe the total mass change in the cylinder, i.e. the net difference between inflowing and outflowing material. Hence, there is not automatically a mass loss when there is some outflowing material. As mentioned before, we only call an outflow a `wind' if it generates mass loss according to this definition.

\C{The middle panel of Fig.~\ref{fig:load_time} shows the mass loss as a function of time. The figure demonstrates that there is no mass loss in Haloes~12.0 and 13.0, despite the outflowing material in Halo~12.0. Hence, with our definition, only Haloes~10.0, 10.5, 11.0 and 11.5 drive a wind, but not Halo~12.0. The mass loss in Halo~11.5 is already intermittent and not continuous in time. For all haloes, the time evolution of the mass loss is similar to the SFR. It is strongest directly after  the onset of the wind and then slowly declines. The absolute value of the mass loss increases with halo mass but seems to level off for halo masses above $10^{10.5}\rmn{~M_\odot}$.}

\C{We remark that the galaxies start losing mass before the outflow reaches the virial radius, in other words before we measure positive outflow velocities. The reason for this effect can be seen  in Fig.~\ref{fig:projections_boxes}. The figure shows projections of the z-component of the velocity for Haloes~10.0, 10.5, 11.0 and 11.5 after $1\rmn{~Gyr}$ (top panel) and $2.5\rmn{~Gyrs}$ (bottom panel). We make these projection in the same way as the projections that are shown in Fig.~\ref{fig:projections}. The displayed region varies from $\pm 40\rmn{~kpc}$ in Halo~10.0 to $\pm 120\rmn{~kpc}$ in Halo~11.5. The black boxes illustrate the sizes of the cylinders that we use to measure mass loss.  The top panel shows that in Haloes~10.0, 10.5 and 11.0 the wind is initially spherical and we start measuring mass loss as soon as this sphere reaches the side of the cylinder. The smaller the halo mass and thus, the radius of the cylinder, the earlier this happens. In Haloes~10.0 and 10.5, the outflow remains mostly spherical whereas a biconical outflow develops in the wake of the spherical component in Halo~11.0. In all haloes, we start measuring outflow velocities when the outflow reaches the top and bottom of the cylinder. In Halo~11.5, the spherical component is much weaker overall and we only start measuring mass loss when the biconical wind reaches the top and bottom of the cylinder. Therefore, the mass loss starts considerably later and also the time delay between mass loss and positive outflow velocity is much smaller. Since the details of the onset of the wind probably depend on our simplified initial conditions, we focus our further analysis on later times when the wind has fully developed. We also keep the small radius of the cylinder to probe the wind-dominated region.}

With the previously determined quantities, we calculate the mass loading of the wind, which is defined as $\dot{M}/\rmn{SFR}$.  We compare mass loss and SFR at the same time and do not model a temporal offset between these quantities. We show the mass loading as a function of time in the fourth panel of Fig.~\ref{fig:load_time}. Remarkably, the mass loading stays almost constant with time in all four haloes, even though SFR and mass loss change.  Furthermore, the mass loading of CR-driven winds is a strong function of halo mass with a value of~$\sim30$ in Halo~10.0 and $2$ in Halo~11.0.  We discuss the mass loading as a function of halo mass in more detail in the next section.

Similar to the mass loading of the wind, we also compare the kinetic energy in the wind with the CR energy that is injected by SN feedback. We obtain the wind energy from the mass loss and the outflow velocity as 
\begin{equation}
\dot{E}_\rmn{w} = \frac{1}{2}\dot{M} \vv_\rmn{out}^2.
\end{equation}
\C{Here, we assume that all mass is lost with the outflow velocity at the virial radius. This is an overestimate for the gas that leaves through the sides of the cylinder, whose velocity can be significantly lower than $\vv_\rmn{out}$. Therefore, our results for the energy loading can only be considered as upper limits.}
The CR energy that is injected per solar mass of star formation is $\varepsilon_\rmn{cr} = 10^{48}\rmn{~erg\;M_\odot^{-1}}$ in the fiducial simulations that we consider here. Hence, the energy loading is given by $\dot{E}_\rmn{w}/(\varepsilon_\rmn{cr} \;\rmn{SFR})$. 

The bottom panel of Fig.~\ref{fig:load_time} shows the energy loading as a function of time. In contrast to the mass loading, the energy loading is neither constant in time nor a strong function of halo mass. 
In Haloes~10.0 and 10.5, the energy loading shows a small peak around $1\rmn{~Gyr}$. It is related to the mass loss before the wind reaches the virial radius and does not represent the wind properties correctly.
The energy loading in Halo~10.0  reaches its maximum directly after the onset of the wind and then decreases with time. In Haloes~10.5, 11.0 and 11.5, the energy loading stays roughly constant and is largest in Halo~11.0. Therefore, no simple scaling with halo mass exists.
\C{Overall, even the upper limits for the energy loading are lower than the mass loading. Typical values range between $1$ and $20$ per cent. Only in Halo~11.0 does the energy loading factor reach unity for more than a Gyr.}


\subsubsection{Scaling of mass and energy loading with halo mass}
\label{sec:massloading}

\begin{figure}
\includegraphics{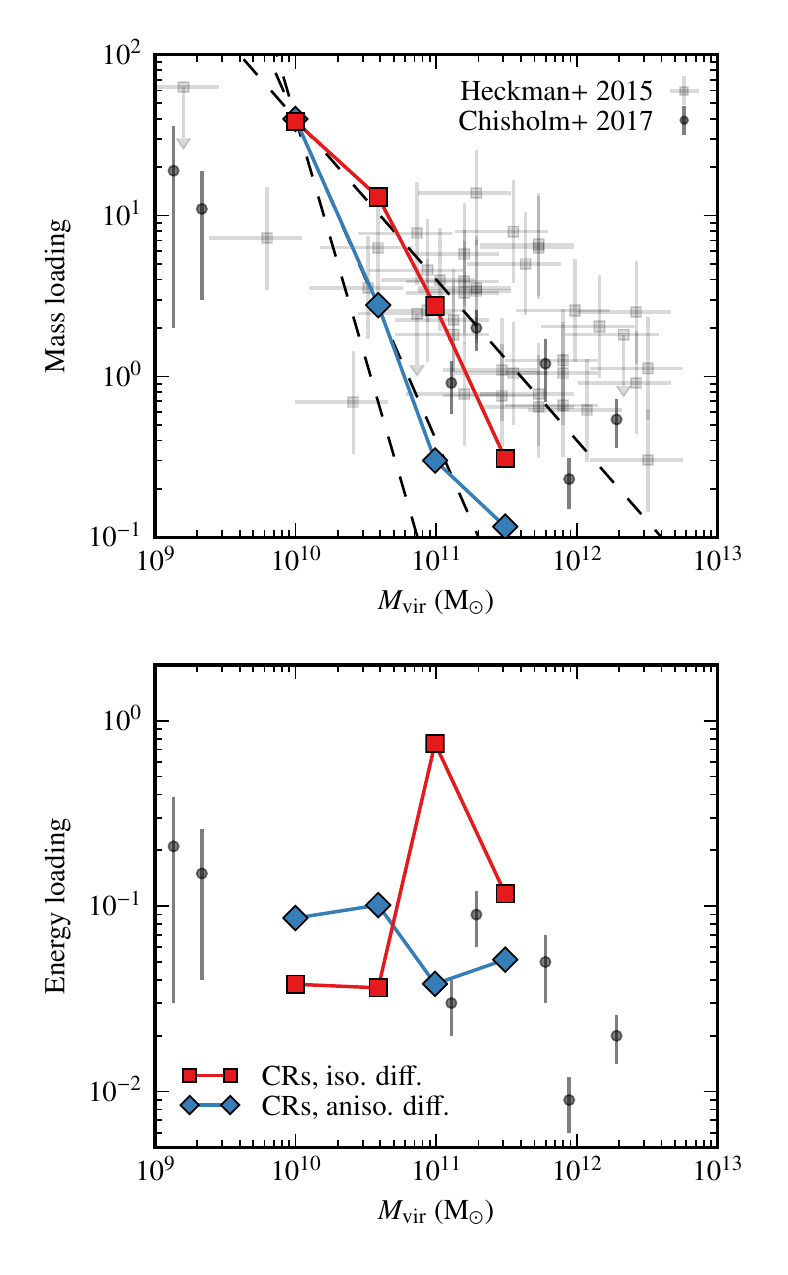}
\caption{The top panel shows the time averaged mass loading factor as a function of halo mass, the bottom panel shows the energy loading factor as a function of halo mass. We average the mass loading between $2$ and $6\rmn{~Gyrs}$ and the energy loading between $3$ and $6\rmn{~Gyrs}$. The dashed lines indicate the power laws $M_\rmn{vir}^{-1}$, $M_\rmn{vir}^{-2}$ and $M_\rmn{vir}^{-3}$. The data points are taken from \citet{Heckman2015} and \citet{Chisholm2017}. The mass loading factors in the simulations of CR-driven winds drop rapidly with halo mass, much faster than in observations. The energy loading does not show a clear scaling with halo mass in our simulations.}
\label{fig:load_mass}
\end{figure}

We are particularly interested in how mass and energy loading scale with halo mass and how this compares to observations. The top panel of Fig.~\ref{fig:load_mass} shows the time averaged mass loading factor as a function of halo mass for the simulations with isotropic (red squares) and anisotropic diffusion (blue diamonds). The bottom panel displays the same for the energy loading. We average between $2$ and $6\rmn{~Gyrs}$ for the mass loading and between  $3$ and $6\rmn{~Gyrs}$ for the energy loading. The reason for the different time intervals is that the energy loading depends on the outflow velocity at $r_\rmn{vir}$, which the outflow reaches only after $\sim 2\rmn{~Gyrs}$.

The figure shows that the mass loading factor of CR-driven winds drops rapidly with halo mass. If we approximate this function with a power law, we obtain a slope that is close to $-2$ for most halo masses. However in some mass ranges, the slope becomes shallower and is closer to $-1$.  For comparison, the dashed lines in Fig.~\ref{fig:load_mass} indicate the power laws $M_\rmn{vir}^{-1}$, $M_\rmn{vir}^{-2}$ and $M_\rmn{vir}^{-3}$.  With the observed scaling, the mass loading of CR-driven winds decreases faster with halo mass than what is expected from purely energy driven winds with a slope of $-2/3$ or purely momentum driven winds with a slope of $-1/3$ \citep{Vogelsberger2013}.  The slopes are rather similar for isotropic and anisotropic diffusion although the mass loading is overall higher for isotropic diffusion.


Next, we compare our results to the observations of nearby starburst galaxies from \citet{Heckman2015} and \citet{Chisholm2017}.\footnote{We assume $\vv_\rmn{vir} \sim \vv_\rmn{circ}$ \citep[see footnote in][]{Heckman2015} and use $\vv_\rmn{vir}$ and $H_\rmn{0} = 100\rmn{~km\;s^{-1}\;Mpc^{-1}}$ (as in the rest of the paper) to calculate $M_\rmn{vir}$.} \C{Both studies use ultraviolet absorption lines to measure outflow velocities and determine mass loss rates with the help of estimates for the geometry and density of the outflow. Since it is extremely challenging to measure mass loss in this way, the inferred mass loading factors are rather uncertain. Moreover, this procedure is very different to the way we measure mass loss in simulations. Similarly, SFRs have to be inferred observationally from infrared and ultraviolet luminosities whereas we can determine it directly from our simulations.} Despite these caveats, we compare our results to the observations in Fig.~\ref{fig:load_mass}.  The figure shows that the overall magnitude of the mass loading of CR-driven winds is in agreement with the observations. However, the mass loading drops much faster with mass in the simulations than in the observations. While \citet{Heckman2015} find no significant scaling with mass, \citet{Chisholm2017} find a weak scaling of $M_\rmn{vir}^{-1/2}$. Both results are much shallower than the scaling of $\sim M_\rmn{vir}^{-2}$ that we find for CR-driven winds.  Furthermore, many winds are observed in galaxies with masses $\gtrsim10^{12}\rmn{~M_\odot}$ in contrast to our simulations. A possible explanation is our definition of mass loss, which requires a net mass loss from a comparatively large cylinder around the galaxy.  Another reason might be that the starbursts in the simulations are not strong or spatially concentrated enough. We explore this possibility in Section~\ref{sec:starburst} but find that the bulk properties of the starbursts are comparable in the simulations and observations.

The bottom panel of Fig.~\ref{fig:load_mass} \C{shows the upper limits for the energy loading} of the simulated CR-driven winds as a function of halo mass. We find no clear scaling with halo mass, neither for isotropic nor for anisotropic diffusion. In most haloes, the energy loading factor reaches values between $3$ and $10$ per cent. \C{Only in Halo~11.0 does the upper limit increases to the remarkably high value of $75$ per cent. The large scatter} without a clear scaling might be in part due to uncertainties in the measurement of the outflow velocity. The comparison of our results to observational data from \citet{Chisholm2017} shows that observed and simulated values are of the same order of magnitude, although the energy loading seems to be a bit low in the least massive halo.

\subsubsection{Impact of model parameters}

\begin{figure}
\includegraphics{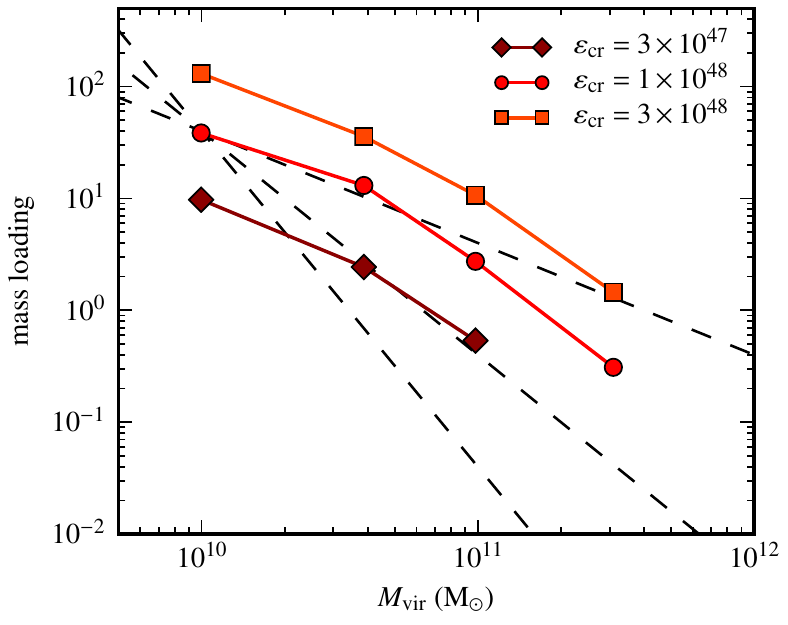}
\caption{Mass loading as a function of virial mass for different CR injection efficiencies, $\varepsilon_\rmn{cr}$ (in $\rmn{erg\;M_\odot^{-1}}$). The dashed lines indicate the power laws $M_\rmn{vir}^{-1}$, $M_\rmn{vir}^{-2}$ and $M_\rmn{vir}^{-3}$. The normalization of the mass loading increases with $\varepsilon_\rmn{cr}$ but the scaling with halo mass remains similar.}
\label{fig:efficiencies}
\end{figure}

\begin{figure}
\includegraphics{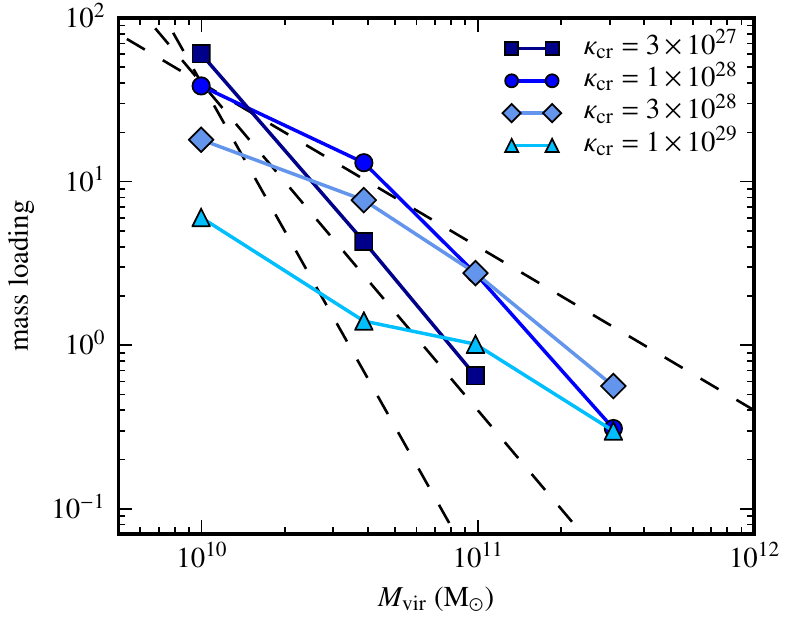}
\caption{Mass loading as a function of virial mass for different CR diffusion coefficients, $\kappa_\rmn{cr}$ (in $\rmn{cm^2\;s^{-1}}$). With increasing $\kappa_\rmn{cr}$, the slope becomes shallower. However, it still remains steeper than $M_\rmn{vir}^{-1}$.
}
\label{fig:kappas}
\end{figure}

The CR model includes parameters that we have not changed so far, but which nevertheless might influence the wind properties.  One important parameter is the CR injection efficiency, $\varepsilon_\rmn{cr}$, which describes the amount of SN energy that is transferred to CRs as they are accelerated at the SN remnant. The CR injection efficiency has a fiducial value of $1\times10^{48}\rmn{~erg}$ per solar mass of star formation. To better understand its effect on the wind, we repeat the simulations with isotropic diffusion of Haloes~10.0, 10.5, 11.0 and 11.5 with both lower and higher values of $\varepsilon_\rmn{cr} = 3 \times 10^{47}\rmn{~erg\;M_\odot^{-1}}$ and $\varepsilon_\rmn{cr} = 3 \times 10^{48}\rmn{~erg\;M_\odot^{-1}}$, respectively.  Analogous to Fig.~\ref{fig:load_mass}, Fig.~\ref{fig:efficiencies} shows the time averaged mass loading factor as a function of halo mass. For all halo masses, the mass loading decreases with decreasing CR efficiency. In the simulation of Halo~11.5 with $\varepsilon_\rmn{cr} = 3 \times 10^{47}\rmn{~erg\;M_\odot^{-1}}$, the wind efficiency is so low that we only detect mass loss in a single snapshot. Thus, we do not show a time averaged mass loading factor for this simulation in Fig.~\ref{fig:efficiencies}.  In contrast to the normalization, the shape of the mass loading as a function of halo mass remains almost the same. For all $\varepsilon_\rmn{cr}$, the mass loading scales as $\sim M_\rmn{vir}^{-1}$ between $10^{10}$ and $10^{11}\rmn{~M_\odot}$. At higher masses, the mass loading drops more rapidly such that it becomes almost proportional to $M_\rmn{vir}^{-2}$.  These results are in agreement with \citet{Salem2014a}, who also find stronger winds with higher mass loading factors if they increase the CR injection efficiency.

A second crucial parameter is the diffusion coefficient.
We study how the mass loading changes  if we vary $\kappa_\rmn{cr}$ in the simulations with isotropic diffusion of Haloes~10.0, 10.5, 11.0 and 11.5.
In addition to our fiducial value of $\kappa_\rmn{cr} = 10^{28}\rmn{~cm^2\;s^{-1}}$, we repeat the simulations with  $\kappa_\rmn{cr} = 3\times10^{27}\rmn{~cm^2\;s^{-1}}$, $\kappa_\rmn{cr} = 3\times10^{28}\rmn{~cm^2\;s^{-1}}$  and  $\kappa_\rmn{cr} = 1\times10^{27}\rmn{~cm^2\;s^{-1}}$.
The time averaged mass loading factors as a function of halo mass are shown in Fig.~\ref{fig:kappas}.
For the lowest $\kappa_\rmn{cr}$, the mass loading drops steeply with virial mass and scales as $M_\rmn{vir}^{-2}$. For this $\kappa_\rmn{cr}$, no continuous wind develops in Halo~11.5 (mass loss only in 7 snapshots between $\sim 5.5$ and $6\rmn{~Gyrs}$).
If the diffusion coefficient is increased, the relation between mass loading and halo mass becomes shallower. For $\kappa_\rmn{cr} = 1 \times 10^{29}\rmn{~cm^2\;s^{-1}}$, the slope is only $-1$. Although there is a systematic trend, the slope remains steep in comparison with the results of \citet{Muratov2015}, who find a slope of $-0.3$, and \citet{Heckman2015}, who find no scaling with halo mass.
For a given halo mass, the mass loading factor does not change with $\kappa_\rmn{cr}$ in a simple way. Since we focus here on the dependence of the wind properties on halo mass, we leave a detailed study of the effects of different diffusion coefficients to future work.


\section{Discussion}
\label{sec:discussion}

In this section, we discuss further aspects of our simulations. We first consider their starburst properties and compare our results to the empirical wind model and other previous works. We then test whether our conclusions are affected by numerical parameters and discuss the limitations of the simulations. 

\subsection{Starburst properties}
\label{sec:starburst}

\begin{figure*}
\includegraphics[width = \textwidth]{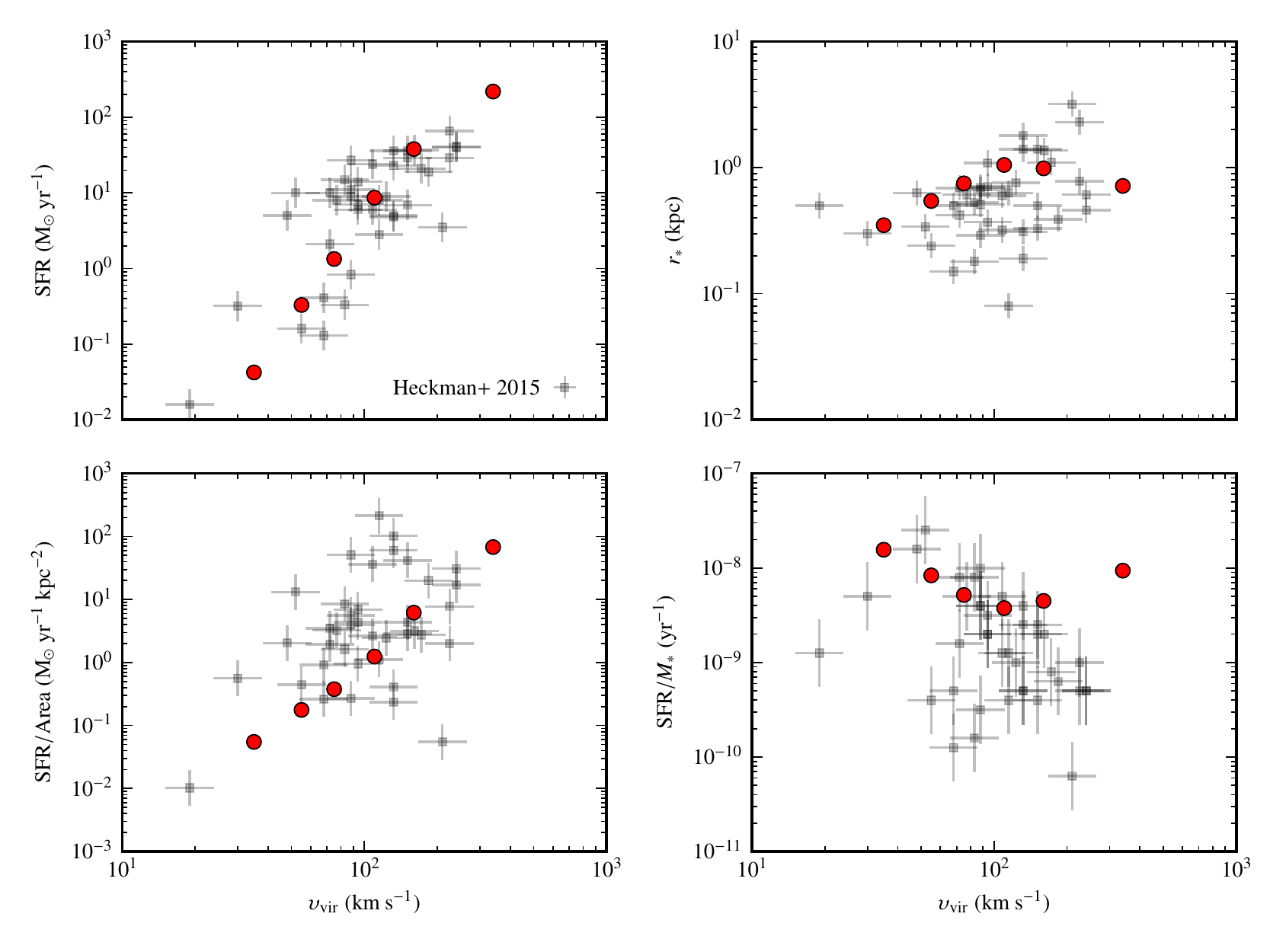}
\caption{Comparison between observed \citep{Heckman2015} and simulated (red dots) starburst properties as a function of virial velocity. We show the results for the simulations with isotropic CR diffusion and measure all properties at the peak of the SFR. The top-left panel shows the SFR and the top-right panel shows a typical radius of the starburst. Observationally, this radius is the half-light radius in the UV images. In the simulations, we use the scale height of the stellar surface density. In the bottom-left panel, the SFR is normalized by the area of the starburst and in the bottom-right panel it is normalized by stellar mass. Overall, the properties of the starbursts in the simulations resemble the observations.}
\label{fig:compdata}
\end{figure*}

In Section~\ref{sec:massloading}, we compared the mass loading that we obtain in our simulations with observations. Here, we study the properties of the starbursts themselves since a stronger starburst likely creates a stronger wind and vice versa. Thus, a comparison of the wind properties is only meaningful if the starburst properties are comparable.

We analyse four different starburst properties as a function of virial velocity, and thus of halo mass, in Fig.~\ref{fig:compdata}. We take the observed circular velocity as an approximation of the virial velocity and assume $\vv_\rmn{circ} \sim \vv_\rmn{vir}$. Moreover, we focus on the properties that are listed for each galaxy in \citet{Heckman2015} due to their larger sample size. The top-left panel of Fig.~\ref{fig:compdata} shows the SFR. The simulated galaxies are characterized by the maximum SFR in the runs with isotropic CR diffusion.  The figure demonstrates that the SFRs in the simulations agree well with observations, although they correlate more strongly with halo mass. Most likely this is a result of the setup of our runs as self-similar, isolated galaxies, which does not allow for realistic variations in the formation history of the galaxies. The SFR in the most massive halo is probably also a bit too high since we do not model AGN feedback, which would additionally reduce the SFR in this mass range.

We show a typical radius of the starburst, $r_\ast$, as a function of virial velocity in the top-right panel of Fig.~\ref{fig:compdata}. In \citet{Heckman2015}, this is the half-light radius of the UV image, which shows the location of recently born stars.  Since the half-light radius is intricate to determine in simulations, we use the scale radius of the stellar surface density instead. We calculate this radius for the snapshot with the maximum SFR. Since the SFR peaks before $0.5\rmn{~Gyr}$ for all halo masses, this snapshot contains mostly young stars. Then, we determine the stellar surface density as a function of the 2D radius in the $xy$-plane and fit an exponential profile to obtain the scale radius. While the exponential profile is a good description in the lower mass haloes, the more massive haloes develop bulges. However, as we only aim for a rough estimate of the typical radius of the starburst, which might also happen in the bulge, we still fit an exponential profile.  Fig.~\ref{fig:compdata} shows that the inferred radii are well in the range of the observed half-light radii. The presence of a bulge in the more massive haloes leads to a decrease of the typical radius.

In the bottom-left panel of Fig.~\ref{fig:compdata}, we normalize the SFR by the area of the starburst, which is given by $2 \pi r_\ast^2$. As before, we show this quantity as a function of virial velocity. The SFR/Area in the simulations is of the same order of magnitude as in the observations. Hence, the spatial concentration of star formation, and thus SN feedback, is comparable.  This might be important since we would expect that it is more difficult to drive outflows if the star formation is more distributed. The SFR/Area in the simulations scales more strongly with virial velocity than in the observations. The reason lies in the rapid increase of the SFR, whereas, in comparison, the radius $r_\ast$ changes only slightly.

We show the gas consumption rate into stars, which is the SFR normalized by stellar mass, as a function of virial velocity in the bottom-right panel of Fig.~\ref{fig:compdata}. In the simulations, we calculate this quantity at the time of the maximum SFR, i.e.~at the peak of the initial starburst. At this early time, not many stars have formed yet such that the simulations have significantly less stellar mass than the observed galaxies. The lack of stellar mass is strongest in the most and least massive haloes, and smallest in haloes with virial velocities between $100$ and $200\rmn{~km\;s^{-1}}$.  Fig.~\ref{fig:compdata} shows that the ratio $\rmn{SFR/}M_\ast$ first decreases with virial velocity, reaches a minimum at roughly $100\rmn{~km\;s^{-1}}$, and then increases again. This dependence on $\vv_\rmn{vir}$ is in parts due to the lack of stellar mass. In addition, we overpredict the SFR in the haloes with the highest masses, which further increases the gas consumption rate. Still, the values for the ratio $\rmn{SFR/}M_\ast$ are of the same order of magnitude as in the observations.

In conclusion, the starburst properties in the simulated and observed galaxies are similar. Hence, their potential for driving winds should also be comparable. The reason for the lack of CR-driven winds in galaxies with halo masses above $\sim 10^{12}\rmn{~M_\odot}$ is probably not due to weaker starbursts. However, our analysis is still fairly rough and we have only looked at a few properties of the starburst. A more careful study that includes mock observations would be necessary to improve the comparison with observations but this goes beyond the scope of the current paper.

\subsection{Comparison with empirical wind model}

In this section, we study in more detail the differences between CR-driven winds and the winds that are launched by the empirical model from \citet{Vogelsberger2013}. 

The wind model creates `wind particles' that are then temporarily decoupled from hydrodynamics until they escape from the star forming phase.  The mass of these particles can be thought of as a `mass loading' at the base of the wind, which we refer to as `particle mass loading'. The particle mass loading scales as $M_\rmn{vir}^{-2/3}$, which corresponds to purely energy driven winds. This scaling is clearly less steep than the $M_\rmn{vir}^{-2}$ dependence that we find for the `global' mass loading of CR-driven winds. Moreover, the model assumes a particle mass loading factor that is roughly an order of magnitude higher than the global mass loading factor of CR-driven winds.

Since the comparison between particle and global mass loading is somewhat ill-defined, we also try to directly compare the global mass loading factors of both wind types. However, the winds that are created by the empirical model have a different morphology than the CR-driven winds. They typically do not have a coherent large scale structure and rarely reach the virial radius. The velocity is overall lower and steadily increases with halo mass. In contrast to CRs, the wind model can still drive outflows in the most massive halo with $10^{13}\rmn{~M_\odot}$. Moreover, the cylinder that we use to measure mass loss has a height equal to the virial radius and is thus not well suited for the winds that are driven by the empirical model. Nevertheless, if we apply it, we measure some mass loss through the sides of the cylinder, which results in mass loading factors between $0.1$ and $1$ with a very weak dependence on halo mass.

\subsection{Comparison with previous work}

We first compare our results to the simulations of CR-driven winds in a $10^{12}\rmn{~M_\odot}$ galaxy from \citet{Salem2014a}. While we find no significant CR-driven outflows at this halo mass, they find strong winds with a mass loading factor of $0.3$.  Their simulations differ in various ways from ours: they are run with an adaptive mesh refinement code, have different subgrid prescriptions for star formation and stellar feedback, and use a different solver for CR diffusion. Moreover, \citet{Salem2014a} start with a pre-existing gas disk and measure mass loading differently. In addition, one of the major differences is that their model neglects CR cooling. In contrast, our model accounts for Coulomb and hadronic losses.
 To test the impact of CR cooling, we re-run the $10^{12}\rmn{~M_\odot}$ halo without cooling. As in \citet{Salem2014a}, we use isotropic CR diffusion. Without CR cooling, a wind develops after $\sim 3\rmn{~Gyrs}$ with a time averaged mass loading factor close to $0.3$. This demonstrates the importance of CR cooling for the properties and occurrence of CR-driven winds. 

 Next, we contrast CR-driven winds with the winds in the Feedback in Realistic Environments (FIRE) simulations \citep{Hopkins2014}, which have been studied by \citet{Muratov2015}.  The FIRE simulations are cosmological zoom simulations of galaxies with masses between $\sim 10^{9}$ and $10^{12}\rmn{~M_\odot}$ at redshift $2$. Stars are formed in bursts that are followed by outflows.  These outflows are driven by a combination of ``early feedback'' from young stars (radiation pressure, stellar winds and ionizing feedback) and energy and momentum input from SNe.  \citet{Muratov2015} find that the mass loading scales with $M_\rmn{vir}^{-1.1}$ if $M_\rmn{vir} < M_\rmn{vir, 60}$ and with $M_\rmn{vir}^{-0.33}$ if $M_\rmn{vir} > M_\rmn{vir, 60}$, where $M_\rmn{vir,60}$ is the halo mass that corresponds to a virial velocity of $60\rmn{~km\;s^{-1}}$.  Both power laws are significantly flatter than what we find for CR-driven winds. Though, at least for $\vv_\rmn{vir} = 75\rmn{~km\;s^{-1}}$ and $z=0.7$ (Halo~11) the mass loading is similar, with a value of $11$ in FIRE and $13$ for CR-driven winds. This comparison could be expanded significantly if we had cosmological zoom simulations of CR-driven winds.


\subsection{Impact of numerical parameters}

\begin{figure*} \includegraphics[width = \textwidth]{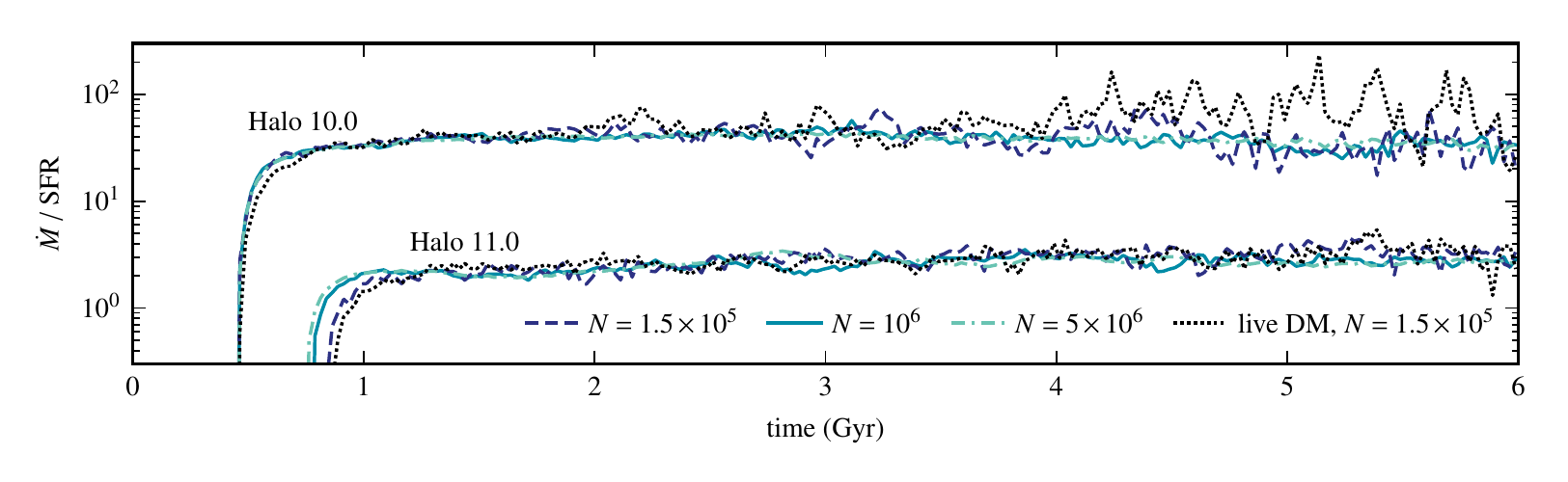} \caption{We vary the resolution and the treatment of dark matter in the simulations with isotropic CR diffusion. Here, we show the resulting time evolution of the mass loading factor for two halo masses. The solid blue lines show our fiducial runs with a static dark matter potential and initially $10^6$ gas cells. The dashed purple lines and the dot-dashed green lines show the simulations with lower and higher resolution, respectively. The black dotted lines give the results if we also simulate the dynamics of the dark matter particles instead of using a stationary dark matter potential. In this case, both, gas and dark matter, are represented by initially $1.5\times10^5$ cells. Overall, the mass loading factor does not change significantly with these variations.}
\label{fig:robust}
\end{figure*}

Our simulations also include numerical parameters whose detailed settings ideally should not have any effect on the results. Here, we study how robust the time evolution of the mass loading factor is if two of these parameters are varied.

First, we analyse the impact of numerical resolution. To this end we re-run the simulations of Haloes~10.0 and 11.0 with isotropic CR diffusion.  In a first test run, we reduce the resolution of our fiducial setup from initially $10^6$ to only $1.5\times10^5$ gas cells. Then, we increase this number to $5\times10^6$. The results are shown in Fig.~\ref{fig:robust}. The dashed, purple line indicates the simulation with lower resolution, the solid, blue line indicates the fiducial simulation and the dot-dashed, green line indicates the simulation with higher resolution. The time evolution of the mass loading factor is fairly similar in all simulations. It is smoother in the runs with higher resolution but the time averaged value is always the same. The results for the energy loading are also rather similar at all resolutions, although the wind velocity at the virial radius in Halo~11.0 changes somewhat.

Furthermore, we study how the representation of the dark matter halo influences the results. In our fiducial setup, we use a static background potential. While this approach is computationally cheap and avoids additional noise from the dark matter halo, it cannot capture back reactions of the baryons on the dark matter potential. Thus, we test whether our results change if we also follow the dynamics of the dark matter particles. Therefore, we represent the gas by $1.5 \times 10^5$ cells and the dark matter with an equal number of particles. Then, we evolve the combined system in time. The dotted, black line in Fig.~\ref{fig:robust} shows the results for Haloes~10.0 and 11.0. The mass loading factor changes slightly in Halo~10.0 but remains essentially the same in Halo~11.0. Thus, a live dark matter halo would not change our overall conclusions.


\subsection{Limitations of the simulations}

An obvious limitation of our simulations is their setup as isolated rotating gas spheres, which cool and form disk galaxies inside-out. Hence, the entire gas supply of the final galaxy collapses at once and creates a huge starburst at the beginning of the simulation. Since star formation is accompanied by CR injection, this might artificially boost the winds. We would need to run fully cosmological (zoom) simulations to study how a more realistic hierarchical assembly history, with sporadic gas accretion and bursty star formation, influences the wind properties. However, this goes beyond the scope of this paper.

Furthermore, there is still some debate about the details of the plasma physics that governs CRs \C{\citep{Zweibel2017, Wiener2017}}.  In particular, CR transport is currently modelled in two different ways. The first possibility, which we also use in this paper, is to describe the transport as diffusion \citep{Booth2012,Hanasz2013,Salem2014a, Pakmor2016wind}. In this approach, CRs diffuse either isotropically or along magnetic field lines with a diffusivity that is close to the Galactic value. The diffusion coefficient is typically kept constant \citep[an exception is e.g.][]{Farber2017}, although it is expected from theory that it changes in space and time \citep{Ptuskin1997, Wiener2013}. A  varying diffusion coefficient might have an impact on the dependence of the mass loading on halo mass. Diffusive CRs do not lose additional energy.  

The second way of describing CR transport is streaming \citep{Uhlig2012, Ruszkowski2017}. Streaming CRs move at roughly the Alfv\'en speed down their own pressure gradient. In contrast to diffusion, the CRs lose energy through this process and heat the thermal gas. \citet{Wiener2017} compare the two different transport mechanisms in the context of CR-driven winds. They find that the winds that are driven by streaming CRs are generally weaker, produce less mass loss and have smaller velocities.  This result makes it unlikely that streaming CRs are more effective in driving outflows, especially in galaxies with masses around $10^{12}\rmn{~M_\odot}$, but only a dedicated study with streaming CRs can ultimately answer this question.

Moreover, diffusive CRs can, in theory, increase their energy in a peculiar way through adiabatic compression. During the early formation of the galaxy, CRs diffuse away from the galaxy while most of the gas is still collapsing. The CRs do not lose energy in the diffusive process but they gain energy when they are subsequently compressed by the infalling gas. This issue has been studied by \citet{Pfrommer2017Letter}. They find that this	cycle leads to a net increase of CR energy in haloes with $10^{12}\rmn{~M_\odot}$. In less massive haloes, the effect becomes weaker and the outwards advection of the wind wins over the inflow after a short initial phase. Since this effect mostly influences haloes that do not drive a mass-loaded wind according to our definition, it should not have a huge impact on the scaling of the wind properties with halo mass.

\section{Conclusions}
\label{sec:conclusion}

Observations demonstrate the ubiquity of galactic winds in starburst galaxies but the physical mechanism that drives these winds is still uncertain. Among other possibilities, CRs are able to drive outflows if CR transport in the form of diffusion or streaming is taken into account. In this paper, we study how the properties of CR-driven winds depend on halo mass.

We simulate a set of isolated galaxies with halo masses between $10^{10}$ and $10^{13}\rmn{~M_\odot}$. We model CRs as an additional fluid and follow the time evolution of the CR energy density. CRs are advected with the gas and can diffuse either isotropically or anisotropically. They are injected as part of the SN feedback and cool through Coulomb and hadronic losses.

We study which galaxies produce CR-driven outflows and focus on the mass and energy loading of the winds (ratios between mass loss and SFR and kinetic wind energy and SFR, respectively).
Our main results are summarized below.
\begin{itemize}
\item
We only obtain CR-driven winds with a mass loss beyond the virial radius in galaxies with halo masses up to $\sim 3 \times 10^{11}\rmn{~M_\odot}$. In galaxies with higher mass, either no outflows exist or the outflows are too weak to cause significant mass loss.
\item
The outflow in the smallest halo with $10^{10}\rmn{~M_\odot}$ is spherical and reaches velocities of $20\rmn{~km\;s^{-1}}$. With increasing halo mass, the winds become biconical and the velocities reach values of up to $200\rmn{~km\;s^{-1}}$.
\item
CR pressure and CR-driven outflows both reduce star formation. However, their combined effect is not sufficient to reproduce the observed stellar mass to halo mass relation in our idealized setup.
\item
The mass loading factor drops rapidly with halo mass with a power-law scaling between $M_\rmn{vir}^{-1}$ and $M_\rmn{vir}^{-2}$, independent of isotropic or anisotropic diffusion. In contrast, the energy loading shows no clear scaling with halo mass.
\item
In comparison to observed, local starburst galaxies, the mass loading drops too rapidly with halo mass. Moreover, winds are frequently observed in galaxies with masses above $3\times10^{11}\rmn{~M_\odot}$ in contrast to our simulations.
However, this comparison has to be considered with caution since crucial quantities, in particular the mass loss, are measured differently in our simulations and the observations.
\item
The CR injection efficiency changes the normalization of the mass loading but has a minor impact on its scaling with halo mass. In contrast, the CR diffusion coefficient affects this scaling: the higher the diffusion coefficient, the shallower the profile becomes.
\item
CR cooling has a significant impact on the development of winds. When we repeat the simulation of the $10^{12}\rmn{~M_\odot}$ halo without the cooling losses for CRs,  an outflow with substantial mass loss develops.
\end{itemize}
These results provide helpful insights into the properties of CR-driven winds and suggest that they are a prime candidate for accounting for much of the feedback needed in low mass galaxies. It remains an interesting question how our results would change if diffusion is replaced by streaming, something that is left for future investigations. In addition, more realistic formation histories of the galaxies, as in cosmological zoom simulations, and a more sophisticated treatment of the multi-phase interstellar medium would further improve our understanding of CR-driven winds.

\section*{Acknowledgements}
We would like to thank the anonymous referee for a constructive report that helped to improve the paper. The authors would like to thank the Klaus Tschira Foundation.  SJ acknowledges funding by the graduate college {\it Astrophysics of cosmological probes of gravity} by Landesgraduiertenakademie Baden-W\"urttemberg. SJ also acknowledges support by the IMPRS for Astronomy and Cosmic Physics at the University of Heidelberg.  RP, CMS and VS acknowledge support through the European Research Council under ERCStG grant EXAGAL-308037.  VS also acknowledges support through subproject EXAMAG of the Priority Programme 1648 ``Software for Exascale Computing' of the German Science Foundation.  CP acknowledges support through the European Research Council under ERC-CoG CRAGSMAN-646955.



\bibliographystyle{mnras}
\bibliography{bib} 



\appendix



\bsp	
\label{lastpage}
\end{document}